\documentclass[aps,pre,twocolumn,superscriptaddress,showpacs]{revtex4}
\usepackage{amssymb}
\usepackage{color}
\usepackage{graphicx}
\usepackage{amsmath}
\usepackage{times}
\usepackage{bm}
\usepackage{float}

\begin{document}

\title{Inducing chimera-synchronization in symmetric complex networks}

\author{Weijie Lin}
\affiliation{Department of Physics, Zhejiang University, Hangzhou 310027, China}
\affiliation{School of Physics and Information Technology, Shaanxi Normal University, Xi'an 710062, China}

\author{Huiyan Li}
\affiliation{School of Science, Beijing University of Posts and Communications, Beijing 100876, China}

\author{Heping Ying}
\affiliation{Department of Physics, Zhejiang University, Hangzhou 310027, China}

\author{Xingang Wang}
\email[Email address: ]{wangxg@snnu.edu.cn}
\affiliation{School of Physics and Information Technology, Shaanxi Normal University, Xi'an 710062, China}

\begin{abstract}
In a recent study of chaos synchronization in symmetric complex networks [Pecora \textit{et al}., Nat. Commun. {\bf 5}, 4079 (2014)], it is found that stable synchronous clusters may coexist with many non-synchronous nodes in the asynchronous regime, resembling the chimera state observed in regular networks of non-locally coupled periodic oscillators. Although of practical significance, this new type of state, namely the chimera-synchronization state, is hardly generated for the general complex networks, due to either the topological instabilities or the weak coupling strength. Here, based on the strategy of pinning coupling, we propose an effective method for inducing chimera-synchronization in symmetric complex network of coupled chaotic oscillators. We are able to argue mathematically that, by pinning a group of nodes satisfying permutation symmetry, there always exits a critical pinning strength beyond which the unstable chimera-synchronization states can be successfully induced. The feasibility and efficiency of the control method are verified by numerical simulations of both artificial and real-world complex networks, with the numerical results well fitted by the theoretical predictions.
\end{abstract}
\date{\today}

\maketitle

\section{INTRODUCTION}
Chimera state refers to the intriguing spatiotemporal pattern in which regions of coherence and incoherence coexist. This peculiar pattern was first observed and analyzed by Kuramoto in simulating the complex Ginzburg-Landau equation with nonlocal couplings \cite{chimera_Kuramoto}, and was later revisited and named chimera state by Abrams and Strogatz \cite{chimera_Abrams}. For their implications to the phenomenon of unihemispheric sleep observed in dolphins and birds \cite{chimera_meaning}, chimera and chimera-like states have been extensively studied over the past decade \cite{chimera_AME,chimera_C,chimera_DRSD,chimera_GAF,chimera_OMY,chimera_SMK,chimera_SR,chimera_SY}.
By a ring of phase oscillators coupled with the cosine kernel, the exact solution of chimera state was obtained in Ref. \cite{chimera_Abrams}. By a minimal model consisting of two interacting populations of oscillators, the stability and bifurcations of chimera state were analyzed in Ref. \cite{chimera_DRSD}. Besides the original model of coupled phase oscillators, chimera states have also been reported in other types of systems, including different oscillating dynamics (e.g., the periodic and chaotic maps, the Stuart-Landau oscillator, and the Hindmarsh-Rose oscillator \cite{maps1,maps2,SL_model,HR}), different coupling functions (e.g., the time-delay and multi-channel couplings \cite{chimera_GAF, FHN}), and different network structures (e.g., the two-dimensional lattices and complex networks \cite{chimera_SY, chimera_complex}). Moreover, as in-depth studies being conducted, many new properties of chimera state have been disclosed, e.g., the Brownian motion of the coherent region \cite{chimera_OMY}, the transient feature of the chimera pattern \cite{transients}, and the existence of multiple coherent regions \cite{chimera_SR,tworegions1}. In experimental studies, chimera states have been observed in chemical, electronic, and optical systems \cite{chemical,electronic,optical}. Recently, the control of chimera states has also been investigated \cite{control_CE, control_OYP, control_JOM}.

In exploring the collective behaviors of coupled chaotic oscillators, an interesting phenomenon is that the oscillators can be self-organized into synchronous clusters, e.g., the cluster (group) synchronization \cite{CS:BAO,CS:Pikovsky,CS:YZHANG,CS:Hasler}. In cluster synchronization, the motions of the oscillators within the same cluster are highly correlated, and are weakly or not correlated if the oscillators belong to different clusters \cite{CS:Pikovsky}. Recently, cluster synchronization in complex networks has also been studied \cite{cs_ZHOUCS,CS_WANGEPL,cs_FU1,cs_FU2,CS:Pecora2014}. In particular, Pecora $\emph{et al}.$ has studied the generation of cluster synchronization in symmetric complex networks, and found the interesting phenomenon of isolated desynchronization. Different from cluster synchronization, in isolated desynchronization a synchronous cluster is emerged on the background of a large number of desynchronized nodes \cite{CS:Pecora2014}. As this state is very similar in form to the chimera state observed in regular networks of non-locally coupled phase oscillators, here we name it the chimera-synchronization state. According to Ref. \cite{CS:Pecora2014}, the stability of the chimera-synchronization state depends on both the network symmetry and coupling strength, making it difficult to be observed in the general networks. Considering the important implications of chimera-synchronization to the functioning and security of many realistic networks, e.g., the power-grid network, it is desirable if stable chimera-synchronization can be generated by some control methods.

In the present work, we propose an effective control method for inducing chimera-synchronization in asynchronous complex networks. Specifically, pinning a set of symmetric nodes in the network by an external controller, we are able to make only the set of pinned nodes synchronized, while keeping the remaining nodes still desynchronized. We shall present our control method in Sec. II, together with a theoretical analysis on the stability of the chimera-synchronization state. In particular, based on the method of eigenvalue analysis, we shall derive mathematically the necessary conditions for generating chimera-synchronization, and give explicitly the formula for the critical pinning strength. In Sec. III, we shall apply the proposed method to different network models, including a small-size artificial network, the Nepal power-grid network, and a large-size complex network. Chimera-synchronization is successfully induced in all these cases, with the critical pinning strengths numerically obtained in good agreement with the theoretical predications. Discussions and conclusion shall be presented in Sec. IV.

\section{control method and theoretical analysis}

Our model of complex network of coupled chaotic oscillators is described by the differential equations
\begin{equation}
\dot{\mathbf{x}}_i=\mathbf{F}(\mathbf{x}_i)+\varepsilon\sum\limits^{N}_{j=1}w_{ij}\mathbf{H}(\mathbf{x}_j),
\end{equation}
with $i,j=1,2,\ldots,N$ the oscillator (node) indices, $\mathbf{x}_i$ the state vector associated with the $i$th oscillator, and $\varepsilon$ the uniform coupling strength. $\dot{\mathbf{x}}=\mathbf{F}(\mathbf{x})$ describes the local dynamics of the oscillators, which is chaotic and, for the sake of simplicity, is set as identical over the network. $\mathbf{H}(\mathbf{x})$ represents the coupling function. The structural connectivity of the network is captured by the coupling matrix $\mathbf{W}=\{w_{ij}\}$, with $w_{ij}>0$ the coupling strength that node $i$ is received from node $j$. If nodes $i$ and $j$ are not directly connected, we set $w_{ij}=0$. The diagonal elements of $\mathbf{W}$ are set as $w_{ii}=-\sum_j w_{ij}$, so as to make $\mathbf{W}$ a Laplacian matrix. This model of linearly coupled nonlinear oscillators has been widely adopted in literature for investigating network synchronization. In particular, the stability of the global synchronization state can be analyzed by the method of master stability function (MSF) \cite{MSF1,MSF2,MSF3}, which shows that the synchronizability of a network is jointly determined by the network structure and nodal dynamics.

We first describe how to identify the set of nodes supporting potentially a synchronous cluster, based on the the information of network symmetries \cite{SYM:Dhys,SYM:Golubitsky,SYM:Russo,CS:Pecora2014}. Let $i$ and $j$ be a pair of nodes in the network whose permutation (exchange) does not change the system dynamical equations [Eq. (1)], we call $(i,j)$ a symmetric pair and characterize it by the permutation symmetry $\textsl{g}_{ij}$. Scanning over all the node-pairs in the network, we are able to identify the whole set of permutation symmetries \{$\textsl{g}_{ij}\}$, which forms the symmetry group $\textsl{G}$. Each symmetry $\textsl{g}$ can be further characterized by a permutation matrix $\mathbf{R}_\textsl{g}$, with $r_{ij}=r_{ij}=1$ if $(i,j)$ is a symmetric pair, and $r_{ij}=0$ otherwise. $\mathbf{R}_\textsl{g}$ is commutative with the coupling matrix, $\mathbf{R}_\textsl{g}\mathbf{W}=\mathbf{W}\mathbf{R}_\textsl{g}$, and, after operating on $\mathbf{W}$, it only exchanges the indices of nodes $i$ and $j$. The set of permutation symmetries for a complex network in general is huge, but can be obtained from $\mathbf{W}$ by the technique of computational group theory \cite{CGT}. Having obtained the symmetry group $\textsl{G}$, we then can partition the network nodes into clusters according to the permutation orbits, i.e., the subset of nodes permuting among one another by the permutation operations are grouped into the same cluster. In such a way, the network nodes will be grouped into $M$ clusters. Assuming that the network initially is staying on the fully desynchronized state (i.e., no synchronization is established between any pair of nodes), our main objective in the present work is to make one of the $M$ clusters synchronized, while, in the meantime, keeping the remaining nodes still desynchronized.

Our method of inducing chimera-synchronization is the following. Firstly, we select from $M$ clusters the one we want to induce synchronization, e.g., the $l$th cluster which contains $n$ nodes. We denote the set of nodes in cluster $l$ as $V_l$, and, by reordering the network nodes, label them with the new indices $i\in [N-n+1,N]$. Then, we pin all oscillators in cluster $l$ by an external controller. The controller has the same local dynamics and coupling function as the oscillators, but are coupled to the oscillators in the one-way fashion (i.e., the oscillators in cluster $l$ are affected by the controller, but not vice versa). Finally, we increase the pinning strength until the desired chimera-synchronization state is generated. With the pinning control, the dynamics of the networked oscillators is governed by the equations
\begin{equation}
\dot{\mathbf{x}}_i=\mathbf{F}(\mathbf{x}_i)+\varepsilon\sum\limits^{N}_{j=1}w_{ij}\mathbf{H}(\mathbf{x}_j)
+\varepsilon\eta\delta_{i}\left[\mathbf{H}(\mathbf{x}_T)-\mathbf{H}(\mathbf{x}_i)\right],
\end{equation}
with $\eta$ the normalized pinning strength, $\mathbf{x}_T$ the sate of the controller, and $\delta$ the delta function: $\delta_{i}=1$ if $i$ $\in$ $V_l$, and $\delta_{i}=0$ otherwise. The controller has the same dynamics as the oscillators, i.e., $\dot{\mathbf{x}}_T=\mathbf{F}(\mathbf{x}_T)$. The specific questions we are interested and going to address are: Can chimera-synchronization be induced by such a control method? and, if yes, what is the necessary pinning strength?

As nodes inside a cluster are commutative with each other, the chimera-synchronization state is naturally a solution of the system equations. That is, if we set the initial conditions of the oscillators in cluster $l$ to be identical, then during the process of system evolution, the states of these oscillators will be always the same. The chimera-synchronization state, however, might be unstable, due to either the weak coupling strength or the network topology \cite{CS:Pecora2014,cs_FU1,cs_FU2}. In the presence of pinning control, the stability of the chimera-synchronization state can be analyzed by the method of eigenvalue analysis, with the details the following. Denote the chimera-synchronization state as $\mathbf{X}=\mathbf{X}^{dsy}\bigoplus\mathbf{X}^{sy}$, with
$\mathbf{X}^{dsy}=[\mathbf{x}_1,\mathbf{x}_2,\ldots,\mathbf{x}_{N-n}]^T$ and $\mathbf{X}^{sy}=[\mathbf{x}_{N-n+1},\mathbf{x}_{N-n+2},\ldots,\mathbf{x}_N]^T$ the state vectors of the desynchronized and synchronized oscillators, respectively, then, according to the definition of chimera-synchronization, we have $\mathbf{x}_{i}=\mathbf{x}^s$ for $i=N-n+1,\ldots,N$, with $\mathbf{x}^s$ the synchronous manifold of the pinned oscillators. Let $\Delta \mathbf{X}=[\delta \mathbf{x}_1,\delta \mathbf{x}_2,\ldots,\delta \mathbf{x}_N]^T$ be the infinitesimal perturbations added on $\mathbf{X}$, then the evolutions of the perturbations are governed by the following variational equations
\begin{equation}
\delta \dot{\mathbf{x}}_i=\mathbf{DF}(\mathbf{x}_i)\delta\mathbf{x}_i+\varepsilon\sum_{j=1}^N c_{ij}\mathbf{DH}(\mathbf{x}_j)\delta \mathbf{x}_j,
\end{equation}
with $\mathbf{DF}(\mathbf{x})$ and $\mathbf{DH}(\mathbf{x})$ the Jacobin matrices, and $\mathbf{C}$ the controlling matrix: $c_{ii}=w_{ii}-\eta$ for $i\in V_l$ (i.e., the set of nodes inside cluster $l$), and $c_{ij}=w_{ij}$ otherwise.

Let $\mathbf{R}$ be the permutation matrix associated to the nodes in cluster $l$ ($r_{ij}=r_{ji}=1$ if $i$ and $j$ belong to $V_l$, $r_{kk}=1$ for $k\notin V_l$, and $r=0$ for other elements) and $\mathbf{T}$ be the transformation matrix of $\mathbf{R}$ (i.e., $\mathbf{T^{-1}RT}$=$\mathbf{R'}$, with $\mathbf{R'}$ the diagonal matrix), then, transforming Eqs. (3) into the mode space of $\mathbf{R}$, we have the new variational equations
\begin{equation}
\delta \dot{\mathbf{y}}_i=\mathbf{DF}(\mathbf{x}_i)\delta\mathbf{y}_i+\varepsilon\sum_{j=1}^N c'_{ij}\mathbf{DH}(\mathbf{x}_j)\delta \mathbf{y}_j,
\end{equation}
where $\Delta \mathbf{Y}=\mathbf{T}^{-1}\mathbf{X}$ and $\mathbf{C}'=\mathbf{T}^{-1}\mathbf{C}\mathbf{T}$. In the mode space, the new controlling matrix $\mathbf{C'}$ has the blocked form
\begin{equation}
\mathbf{C'}=\left(
  \begin{array}{cc}
    \mathbf{B} & 0 \\
     0    &  \mathbf{D} \\
  \end{array}
\right),
\end{equation}
with $\mathbf{B}$ and $\mathbf{D}$ the $(n-1)$- and $(N-n+1)$-dimensional matrices, respectively. The matrix $\mathbf{B}$ characterizes the perturbations transverse to the synchronous manifold of cluster $l$, we thus call the space it spans  the transverse subspace. (Please note that as $\mathbf{C'}$ and $\mathbf{C}$ are similar matrices, they have the same set of eigenvalues. The significance of the transformation lies in separating the eigenvalues into two different groups.) As the transverse modes are decoupled from the other modes, the synchronizability of the pinned cluster therefore can be analyzed separately. Focusing on only the transverse modes, we have the variational equations
\begin{equation}
\delta \dot{\mathbf{y}}_{i'}=\mathbf{DF}(\mathbf{x}^s)\delta\mathbf{y}_{i'}+\varepsilon\sum_{j'=1}^{n-1} b_{i'j'}\mathbf{DH}(\mathbf{x}^s)\delta \mathbf{y}_{i'},
\end{equation}
with $i',j'=1,2,\ldots,n-1$, $\mathbf{B}=\{b_{i'j'}\}$, and $\mathbf{x}^s$ the synchronous manifold of the pinned cluster.

To make the cluster synchronizable, it is required that $\delta \mathbf{y}_i$ is damping to 0 with time for all the transverse modes -- a question that can be addressed by the MSF method \cite{MSF1,MSF2,MSF3}. To be specific, transforming Eqs. (6) into the new mode space spanned by the eigenvectors of matrix $\mathbf{B}$, we can obtain the decoupled variational equations
\begin{equation}
\delta \dot{\mathbf{z}}_{i'}=[\mathbf{DF}(\mathbf{x}^s)+\varepsilon\lambda_{i'}\mathbf{DH}(\mathbf{x}^s)]\delta \mathbf{z}_{i'},
\end{equation}
where $0>\lambda_1\geq\lambda_2\geq\ldots\geq\lambda_{n-1}$ are the eigenvalues of $\mathbf{B}$, and $\delta\mathbf{z}_{i'}$ is the $i'$th perturbation mode in the new space. To make the cluster synchronization stable, the necessary conditions now become that $\delta\mathbf{z}_{i'}$ should be damping to $0$ with time. Let $\Lambda_{i'}$ be the largest Lyapunov exponent calculated from Eq. (7), then whether $\delta \mathbf{z}_{i'}$ is damping with time is determined by the sign of $\Lambda_{i'}$: the mode is stable if $\Lambda_{i'}<0$, and is unstable if $\Lambda_{i'}>0$. Defining $\sigma\equiv-\varepsilon\lambda$, by solving Eq. (7) numerically we can obtain the function $\Lambda=\Lambda(\sigma)$, i.e., the MSF curve. Previous studies of MSF have shown that for the typical nonlinear oscillators, $\Lambda$ is negative when $\sigma$ is larger than a critical threshold $\sigma_c$, with $\sigma_c>0$ a parameter dependent of both the oscillator dynamics and coupling function. Hence, to keep the pinned cluster synchronizable, it is required that $\sigma_{i'}>\sigma_c$ for all the transverse modes. Since $\lambda_1\geq\lambda_2\ldots\geq\lambda_{n-1}$, this requirement thus can be simplified as
\begin{equation}
\varepsilon|\lambda_1|>\sigma_c.
\end{equation}

Since $\mathbf{B}$ is derived from $\mathbf{C}$ and $\mathbf{C}$ is dependent of both the network coupling matrix, $\mathbf{W}$, and the pinning strength, $\eta$, $\lambda_1$ thus is determined jointly by $\mathbf{W}$ and $\eta$. To have the formula for the critical pinning strength, we need to express $\lambda_1$ as a function of $\eta$ explicitly. Noticing that the controlling matrix $\mathbf{C}$ is constructed from the coupling matrix $\mathbf{W}$ by replacing $w_{ii}$ with $w_{ii}-\eta$ for only the pinned oscillators, $\mathbf{W}$ thus can be also transformed to the blocked form shown in Eq. (5) by the transformation matrix $\mathbf{T}$, $\mathbf{W}'=\mathbf{T}^{-1}\mathbf{W}\mathbf{T}$. Denoting $\mathbf{B}^w$ as the transverse subspace of $\mathbf{W}'$ and let $0>\lambda^w_1\geq\lambda^w_2\geq\ldots\geq\lambda^w_{n-1}$ be the eigenvalues of $\mathbf{B}^w$, it is straightforward to find that $\lambda_{i'}=\lambda^w_{i'}-\eta$ for $i'=1,2,\ldots,n-1$. In particular, we have $\lambda_{1}=\lambda^w_{1}-\eta$ for the first transverse mode, which, inserting into Eq. (8), gives the following formula of the critical pinning strength
\begin{equation}
\eta_c=\sigma_c/\varepsilon-|\lambda^w_{1}|.
\end{equation}

\section{applications}

We next verify the feasibility and efficiency of the proposed control method by applying it to different complex network models, including a small-size network, the Nepal power-grid network, and a large-size complex network.

\subsection{Small-size network}

\begin{figure*}
\begin{center}
\includegraphics[width=0.9\linewidth]{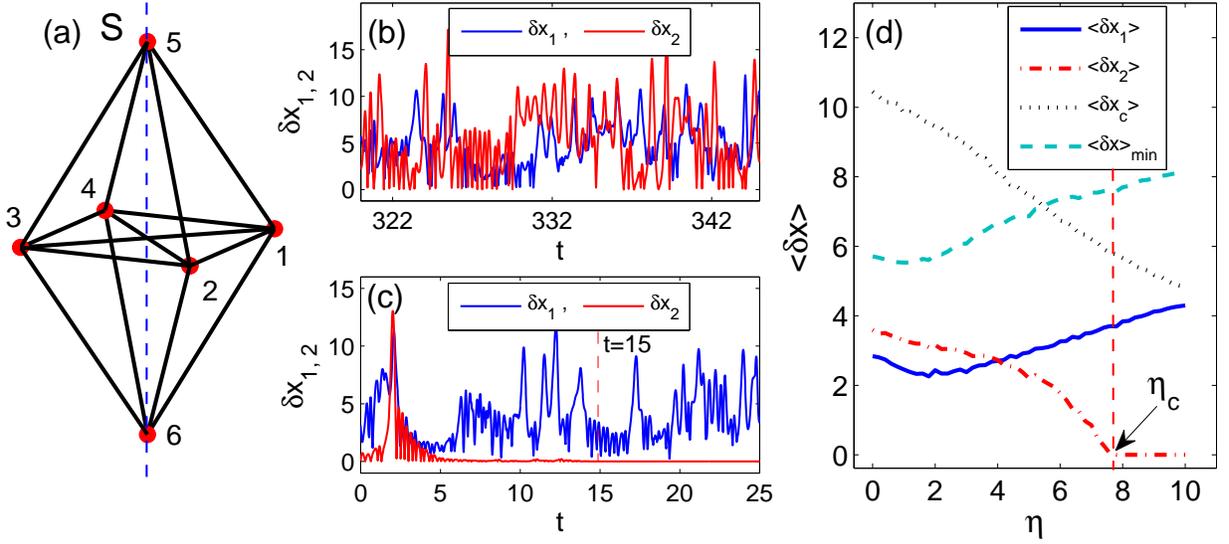}
\caption{(Color online) Inducing chimera synchronization in a six-node network of coupled chaotic Lorenz oscillators. (a) The network structure. Nodes are grouped into two clusters according to their symmetries: $V_1=\{1,2,3,4\}$ (rotation symmetry) and $V_2=\{5,6\}$ (reflection symmetry). (b) By the coupling strength $\varepsilon=0.7$, the time evolution of the cluster-synchronization errors $\delta x_{1,2}$ in the absence of the pinning control. Both cluster are asynchronous. (c) With the pinning strength $\eta=8.0$, the time evolutions of $\delta x_{1,2}$. Cluster 2 is synchronized at about $t=15$, while cluster 1 remains asynchronous throughout the process. (d) The variation of the time-averaged cluster-synchronization errors, $\langle\delta x_{1,2}\rangle$, as a function of $\eta$. $\langle\delta x_2 \rangle\approx 0$ at $\eta_c\approx 7.8$. $\langle\delta x_c\rangle$ is the time-averaged synchronization error between oscillator $5$ and the controller; $\langle\delta x\rangle_{min}$ is the smallest synchronization error between oscillator $5$ and oscillators in cluster 1.} \label{Fig_1}
\end{center}
\end{figure*}

We first demonstrate how to induce chimera synchronization in a small-size network. The structure of the network is presented in Fig. 1(a), which is constructed by deleting one link (e.g., $c_{56}=0$) from a globally connected network of $6$ nodes. For the sake of simplicity, we treat the network links as non-weighted and non-directed, e.g., $w_{ij}=w_{ji}=1$ if there is a link between $i$ and $j$. In simulations, we adopt the chaotic Lorenz oscillator as the nodal dynamics, which in its isolated form is described by the equations $(dx/dt, dy/dt, dz/dt)^{T}=[\alpha(y-x),rx-y-xz,xy-bz]^{T}$. The parameters are chosen as $\alpha=10, r=35$, and $b=8/3$, with which the isolated oscillator presents the chaotic motion, with the largest Lyapunov exponent is about 1.05. The coupling function is chosen as $\mathbf{H}([x,y,z]^T)=[0,x,0]^T$. Having fixed the nodal dynamics and coupling function, we can obtain the MSF curve $\Lambda=\Lambda(\sigma)$ by solving Eq. (7) numerically, which shows that $\Lambda$ is negative in the region  $\sigma>\sigma_c\approx 8.3$ \cite{MSF3}.

The symmetries of the network can be discerned by visual inspection: the group of nodes $(1,2,3,4)$ are of rotation symmetry, and the pair of nodes $(5,6)$ are of reflection symmetry. Accordingly, the nodes can be divided into two clusters: $V_1=\{1,2,3,4\}$ and $V_2=\{5,6\}$. To measure the synchronization degree of the clusters, we introduce the cluster-synchronization error $\delta x_l=\sum_{i=1}^{n_l}{|x_i- \bar{x}_l|}/n_l$, with $i\in V_l$, $n_l$ the cluster size, and $\bar{x}_l=\sum_i x_i/n_l$ the average state of cluster $l$. Clearly, the smaller is $\delta x_l$, the better is the oscillators in cluster $l$ synchronized. Setting $\varepsilon=0.7$, we plot in Fig. 1(b) the evolutions of $\delta x_1$ and $\delta x_2$ with time. It is evident that neither of the clusters is synchronized. To illustrate, we pin oscillators $5$ and $6$ by an external controller according to Eq. (2), so as to induce the synchronization for cluster $2$. (It is worth noting that due to the network topology, the $2$nd cluster can not be synchronized by varying the coupling strength, i.e., it is topologically unstable. In contrast, the $1$st cluster can be synchronized by a larger coupling strength, i.e., it is dynamically unstable.) By the pinning strength $\eta=8.0$, in Fig. 1(c) we plot again the time evolutions of $\delta x_1$ and $\delta x_2$. It is seen that after a transient period about $t=15$, we have $\delta x_2\approx 0$, while $\delta x_1$ is still of large value. Indeed, with the pinning control, the desired chimera-synchronization state can be induced from the asynchronous network.

To find out the critical pinning strength, $\eta_c$, for inducing the chimera-synchronization state, we plot in Fig. 1(d) the variation of the time-averaged cluster-synchronization error, $\left<\delta x_l\right>$, as a function of $\eta$ (the error is averaged over a period of length $t=50$). It is seen that $\left<\delta x_2\right>$ reaches $0$ at about $7.8$, while $\left<\delta x_1\right>$ remains large. We thus have $\eta_c\approx 7.8$. To check whether synchronization is established between the controller and the pinned oscillators, we plot in Fig. 1(d) also the variation of the time-averaged synchronization error between oscillator $5$ and the controller, $\left<\delta x_c\right>=\left<|x_5-x_T|\right>$, as a function of $\eta$. It is seen that $\left<\delta x_c\right>$ remains large when $\eta>\eta_c$, indicating that the synchronous cluster is induced, but not controlled by the external controller. (In our simulations, we have increased $\eta$ up to $30$, and found that the value of $\left<\delta x_c\right>$ is still large.) Meanwhile, to check whether there are other synchronous clusters formed on the network, we plot in Fig. 1(d) also the variation of the smallest synchronization error between oscillators in cluster $1$ and oscillator $5$, $\left<\delta x\right>_{min}=\min\{\left<|x_5-x_j|\right>\}$ with $j\in V_1$. As $\left<\delta x\right>_{min}>0$ in the region $\eta>\eta_c$, the possibility of forming other synchronous clusters thus is excluded.

The critical pinning strength can be analyzed by the method of eigenvalue analysis presented in Sec. II. As nodes $5$ and $6$ are of reflection symmetry, their permutation does not change the system dynamics. We therefore have the permutation matrix
\begin{equation}
\mathbf{R}=\left(
  \begin{array}{cccccc}
    1 & 0 & 0 & 0 & 0 & 0\\
    0 & 1 & 0 & 0 & 0 & 0\\
    0 & 0 & 1 & 0 & 0 & 0\\
    0 & 0 & 0 & 1 & 0 & 0\\
    0 & 0 & 0 & 0 & 1 & 1\\
    0 & 0 & 0 & 0 & 1 & 1\\
  \end{array}
\right),
\end{equation}
from which we can obtain the transformation matrix (constructed by the eigenvectors of $\mathbf{R}$)
\begin{equation}
\mathbf{T}=\left(
  \begin{array}{cccccc}
          0  &   0  &  0  &   0  &  1 &     0\\
          0  &   1  &  0  &   0  &  0 &     0\\
          0  &   0  &  1  &   0  &  0 &     0\\
          0  &   0  &  0  &   1  &  0 &     0\\
-\sqrt{2}/2  &   0  &  0  &   0  &  0 &  \sqrt{2}/2\\
 \sqrt{2}/2  &   0  &  0  &   0  &  0 &  \sqrt{2}/2\\
  \end{array}
\right).
\end{equation}
As nodes $5$ and $6$ and pinned, we have the controlling matrix
\begin{equation}
\mathbf{C}=\left(
  \begin{array}{cccccc}
    -5 &  1 &  1 &  1 &  1 &  1 \\
     1 & -5 &  1 &  1 &  1 &  1 \\
     1 &  1 & -5 &  1 &  1 &  1 \\
     1 &  1 &  1 & -5 &  1 &  1 \\
     1 &  1 &  1 &  1 & -4-\eta &  0 \\
     1 &  1 &  1 &  1 &  0 & -4-\eta \\
  \end{array}
\right),
\end{equation}
which, after the transformation operation $\mathbf{C'}=\mathbf{T}^{-1}\mathbf{C}\mathbf{T}$, has the blocked form show in Eq. (5), with
\begin{equation}
\mathbf{D}=\left(
  \begin{array}{ccccc}
      -5 &      1  &       1 &    1     & \sqrt{2} \\
       1 &     -5  &       1 &    1     & \sqrt{2} \\
       1 &      1  &      -5 &    1     & \sqrt{2} \\
       1 &      1  &       1 &   -5     & \sqrt{2}\\
\sqrt{2} & \sqrt{2}& \sqrt{2}& \sqrt{2} &    -4-\eta  \\
  \end{array}
\right),
\end{equation}
and
\begin{equation}
\mathbf{B}=-4-\eta.
\end{equation}
We thus have $\lambda_1=\lambda^w_{1}-\eta=-4-\eta$, which, according to Eq. (8), gives $\eta_c=\sigma_c/\varepsilon-|\lambda^w_{1}|=8.3/0.7-4\approx7.86$. This prediction is in a good agreement with the numerical result (numerically we have $\eta_c\approx 7.8$).

\subsection{Power-grid network}

\begin{figure}
\begin{center}
\includegraphics[width=0.9\linewidth]{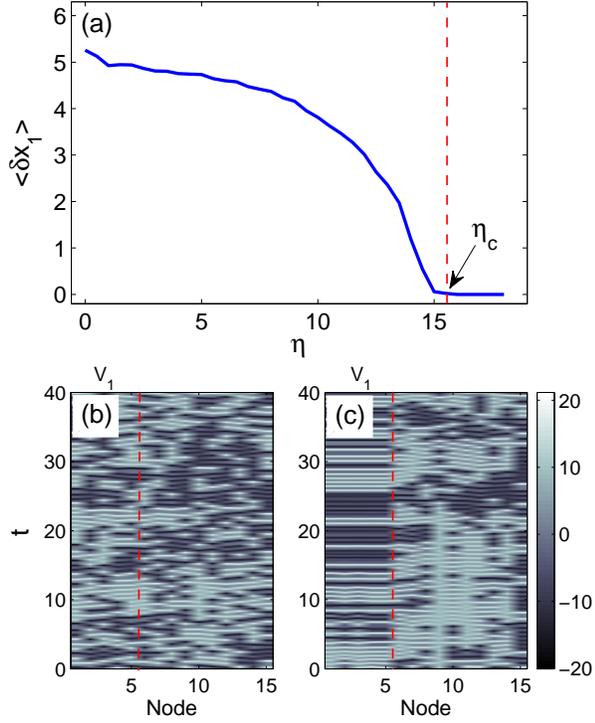}
\caption{(Color online) Inducing chimera synchronization in the network of Nepal power-grid. The nodal dynamics and coupling function are the same to Fig. 1, and the coupling strength is fixed as $\varepsilon=0.32$. (a) The variation of the time-averaged synchronization error of cluster 1, $\langle\delta x_1\rangle$, as a function of the pinning strength, $\eta$. Chimera synchronization is induced when $\eta>\eta_c\approx16$. The spatiotemporal evolution of the oscillators under the pinning strengths (b) $\eta=14$ and (c) $\eta=18$.} \label{Fig_2}
\end{center}
\end{figure}

We next demonstrate how to induce chimera synchronization in a realistic complex network. The network model employed here is the Nepal power-grid \cite{Nepal}, which contains $N=15$ nodes (power stations) and $62$ links (power lines). For the sake of simplicity, we treat the links as non-weighted and non-directed, e.g., $w=1$ for all the network links. By the technique of computational group theory \cite{CS:Pecora2014,CGT}, we are able to figure out all the network permutation symmetries (totally $86400$), and, according to the permutation orbits, partition the nodes into $5$ clusters: $V_1=\{1,2,3,4,5\}$, $V_2=\{6,7,8\}$, $V_3=\{9,10,11,12,13\}$, $V_4=\{14\}$, and $V_5=\{15\}$ \cite{CS:Pecora2014}. Among them, the $4$th and $5$th clusters are trivial, as each contains only a single node. Still, we adopt the chaotic Lorenz oscillator as the nodal dynamics, and use $\mathbf{H([x,y,z]^T)}=[0,x,0]^T$ as the coupling function. The coupling strength is fixed as $\varepsilon=0.32$, with which no synchronization relationship is established between any pair of oscillators on the network.

For illustration purpose, we pin oscillators in cluster $1$ according to Eq. (2). Based on numerical simulations, we plot in Fig. 2(a) the variation of the time-averaged synchronization error of cluster 1, $\langle\delta x_1\rangle$, as a function of the pinning strength, $\eta$. It is seen that $\langle \delta x_1\rangle$ reaches 0 at about $\eta_c\approx 16$. To have a clearer picture on the transition of the system dynamics from the asynchronous to chimera-synchronization states around $\eta_c$, we plot in Figs. 2(b) and (c) the spatiotemporal evolution of the network for different values of $\eta$. For the case of $\eta=14<\eta_c$ [Fig. 2(b)], it is seen that the evolution is random and irregular. For the case of $\eta=18>\eta_c$ [Fig. 2(c)], it is seen clearly that after a transient period about $t\approx 18$, the oscillators in cluster 1 are well synchronized, while the motions of the other oscillators in the network remain uncorrelated.

Still, the critical pinning strength shown in Fig. 2(a) can be analyzed by the method of eigenvalue analysis presented in Sec. II. To save the space, here we omit the detail deduction, but present only the main results. In constructing the permutation matrix $\mathbf{R}$, we set $r_{ij}=r_{ji}=1$ for $i,j\in V_1$, $r_{kk}=1$ for $k\notin V_1$, and $r=0$ for the remaining elements. By the eigenvectors of $\mathbf{R}$, we can construct the transformation matrix, $\mathbf{T}$, and then used it to transform the control matrix $\mathbf{C}$ into the blocked matrix $\mathbf{C}'$ [which has the form shown in Eq. (5)]. From $\mathbf{C}'$, we can obtain the transverse matrix $\mathbf{B}$, which is $4$-dimensional and has the degenerated eigenvalues $\lambda_{1,2,3,4}=\lambda^w_{1}-\eta=-8-\eta$. According to the Eq. (8), we thus have the critical pinning strength $\eta_c=\sigma_c/\varepsilon-|\lambda^w_{1}|=8.3/0.32-8\approx 18$, which agrees with the numerical result very well (numerically we have $\eta_c\approx 16$).

\subsection{Large-size complex network}

\begin{figure}
\begin{center}
\includegraphics[width=0.9\linewidth]{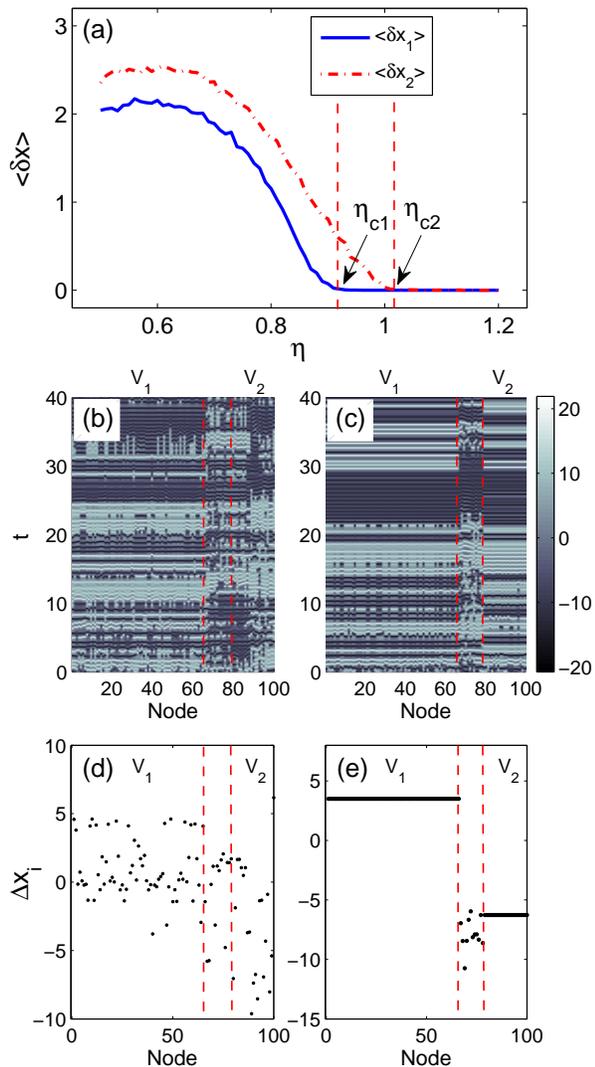}
\caption{(Color online) Inducing two-cluster chimera-synchronization state in a complex network of $N=100$ coupled chaotic Lorenz oscillators. The coupling strength is fixed as $\varepsilon=4.4$. (a) The variation of the time-averaged cluster-synchronization errors, $\langle\delta x_{1,2}\rangle$, as a function of the pinning strength, $\eta$. $\langle\delta x_{1}\rangle$ and $\langle\delta x_{2}\rangle$ reach 0 at $\eta_{c1}\approx0.9$ and $\eta_{c2}\approx1.01$, respectively. The time evolution of the network for $\eta=0.7$ (b) and $\eta=1.15$ (c). The snapshots of the network at $t=20$ for $\eta=0.7$ (d) and $\eta=1.15$ (e). $\Delta x_i=x_i-\overline{x}$, with $\overline{x}=\sum_i x_i/N$ the average state of the network.} \label{Fig_3}
\end{center}
\end{figure}

We finally demonstrate how to induce chimera synchronization in a large-size complex network. Recently, a new type of chimera-state consisting of two or more coherent regions, namely the multiple-cluster chimera state, has been reported in regular networks of coupled periodic oscillators \cite{chimera_SR,tworegions1}. It is intriguing to see whether the similar state can be induced on complex network by the proposed pinning method. To investigate, we generate a random network of $N=100$ nodes and $4931$ links (generated by removing randomly $19$ links from the globally connected network). Again, we adopt the chaotic Lorenz oscillator as the nodal dynamics and use $\mathbf{H([x,y,z]^T)}=[0,x,0]^T$ as the coupling function. This time, to avoid the overflow in numerical simulations, we adopt the normalized coupling scheme $w_{ij}=a_{ij}/k_i$ \cite{NETSYN:MOTTER2005,NETSYN:XGW2007}, with $\mathbf{A}=\{a_{ij}\}$ the adjacency matrix and $k_i=\sum_j a_{ij}$ the node degree (the number of connections for node $i$). In general, we have $w_{ij}\neq w_{ji}$, i.e., the couplings are weighted and directed. By the technique of computational group theory, we are able to identify all the network symmetries, based on which the network nodes are grouped into $4$ clusters. In particular, the largest cluster contains $66$ nodes ($V_1=\{1,2,\ldots,66\}$), and the second largest cluster contains $22$ nodes ($V_2=\{79,80,\ldots,100\}$). We fix the coupling strength as $\varepsilon=4.4$, with which no synchronization is observed between any two oscillators.

To implement the control, we introduce two independent external controllers, $\mathbf{x}_{T1}$ and $\mathbf{x}_{T2}$, with controllers 1 and 2 pin clusters 1 and 2, respectively. The two controllers are of the same dynamics and pinning strength, but are evolving independently. By numerical simulations, we plot in Fig. 3(a) the variation of the time-averaged cluster-synchronization errors, $\langle\delta x_{1,2}\rangle$, as a function of $\eta$. It is seen that $\langle\delta x_{1}\rangle$ and $\langle\delta x_{2}\rangle$ reach $0$ at about $\eta_{c1}\approx 0.9$ and $\eta_{c2}\approx 1.01$, respectively. Therefore, in the region $\eta\in (\eta_{c1},\eta_{c2})$ the network is staying on the one-cluster chimera-synchronization state (only cluster 1 is synchronized), and in the region $\eta>\eta_{c2}$ the network is staying on the two-cluster chimera-synchronization state (both two pinned clusters are synchronized).

To have more details on the formation of the two-cluster chimera-synchronization state, we plot in Figs. 3(b) and (c) the spatiotemporal evolution of the oscillators under different pinning strength. For a weak pinning strength $\eta=0.7<\eta_{c1}$ [Fig. 3(b)], it is seen that the oscillators are evolving independently, without any sign of synchronized clusters. For a strong pinning strength $\eta=1.15>\eta_{c2}$ [Fig. 3(c)], it is seen clearly that the motions of the oscillators in each cluster, $V_{1,2}$, are highly correlated. To characterize feature the chimera-synchronization state, we plot in Figs. 3(d) and (e) the snapshots of the network at the moment $t=20$ of the system evolutions shown in Figs. 3(b) and (c). For $\eta=0.7$ [Fig. 3(d)], it is seen that the states of the oscillators are scattered randomly over a wide range; while for $\eta=1.15$ [Fig. 3(e)], the oscillators in $V_1$ ($V_2$) are of the identical state.

Despite the weighted coupling matrix and the two-pinned clusters, the critical pinning strengths, $\eta_{c1}$ and $\eta_{c2}$, can still be analyzed by the method of eigenvalue analysis. In constructing the permutation matrix $\mathbf{R}$, we set $r_{ij}=r_{ji}=1$ if nodes $i$ and $j$ belong to $V_1$ ($V_2$), and $r_{kk}=1$ for nodes which do not belong to $V_{1}$ and $V_2$. Transformed into the mode space of $\mathbf{R}$, the controlling matrix has the blocked form shown in Eq. (5). Different from the one-cluster case, here there are two transverse subspaces, $\mathbf{B}_1$ and $\mathbf{B}_2$. $\mathbf{B}_1$ is $65$-dimensional, which characterizes the transverse subspace of cluster $1$. The largest eigenvalues of $\mathbf{B}_1$ is $\lambda_{1}=\lambda^w_{1}-\eta=-1.01-\eta$, which, according to Eq. (8), gives $\eta_{c1}=\sigma_c/\varepsilon-|\lambda^w_{1}|=8.3/4.4-1.01\approx0.88$. $\mathbf{B}_2$ is $21$-dimensional, which characterizes the transverse subspace of cluster $2$. The largest eigenvalues of $\mathbf{B}_2$ is $\lambda_{1}=\lambda^w_{1}-\eta=-1.0-\eta$, which, according to Eq. (8), gives $\eta_{c2}=\sigma_c/\varepsilon-|\lambda^w_{1}|=8.3/4.4-1.0\approx0.89$. The theoretical predictions fit the numerical results reasonably well (numerically we have $\eta_{c1}\approx0.9$ and $\eta_{c2}\approx 1.01$). In particular, the theory well predicts that cluster 1 is synchronized by a smaller pinning strentrh as compared to cluster 2.

\section{Discussions and Conclusion}

The proposed method of inducing chimera synchronization can be applied to the general complex networks. In our simulations, we have applied this method to a variety of complex networks (including all the network models studied in Ref. \cite{CS:Pecora2014})), and found that, given the network contains a symmetric cluster, there always exits a critical pinning strength beyond which the desired chimera-synchronization state can be stably generated. Meanwhile, as the underlying mechanism of chimera synchronization is governed by cluster synchronization, the proposed control method can be applied to the general nodal dynamics and coupling functions. For instance, replacing the coupling function with $\mathbf{H}[x,y,z]^T=[x,0,0]^T$ (different from the one demonstrated above, this coupling function generates a bounded stable region in the MSF curve), we have observed the similar chimera-synchronization states shown in Figs. 1-3. Besides the chaotic Lorenz oscillators, we have also tested the other nodal dynamics, including the chaotic R\"{o}ssler and Hindmarsh-Rose oscillators, where the similar chimera-synchronization states can be also successfully induced by proper pinning strengths.

It should be emphasized that the proposed pinning method is able to induce, but not control the synchronous cluster. This property is rooted in the symmetry of the enlarged pinning network, i.e., considering the controller as an additional node to the original network. In this enlarged network, the oscillators in the pinned cluster are still satisfying the permutation symmetry, but they are not exchangeable with the controller. As the pinned oscillators are perturbed by the desynchronized oscillators while the controller is not, it is therefore impossible to make the pinned oscillators synchronize with the controller. However, if the whole network is synchronized (instead of chimera synchronization), it would be possible to control the synchronous manifold. In such a case, the enlarged network will reach the state of global synchronization instead of chimera synchronization \cite{PIN-1}.

To summarize, we have proposed a general pinning method for inducing chimera synchronization in symmetric complex networks of coupled chaotic oscillators, and found that, given the network contains a group of symmetric nodes, there always exits a critical pinning strength beyond which a stable synchronous cluster can be generated on the background of desynchronized nodes. We have conducted a detail analysis on the stability of the chimera-synchronization state, and obtained the formula of the critical pinning strength. The feasibility and efficiency of the control method have been verified by numerical simulations on various network models, with the numerical results in good agreement with the theoretical predictions. Our studies shed new lights on the collective dynamics of complex networks, and might potentially be used to the design of modern control techniques.

This work was supported by the National Natural Science Foundation of China under the Grant No.~11375109 and by the Fundamental Research Funds for the Central Universities under the Grant No.~GK201303002.


\begin{thebibliography}{99}

%%%%%%%%%%%%%  chimera literature
\bibitem{chimera_Kuramoto} Y. Kuramoto and D. Battogtokh, Coexistence of coherence and incoherence in nonlocally coupled phase oscillators, Nonlinear Phenom. Complex Syst. {\bf5}, 380 (2002).   %%% first observed

\bibitem{chimera_Abrams} D. Abrams and S. Strogatz, Chimera states for coupled oscillators, Phys. Rev. Lett. {\bf93}, 174102 (2004).  %%%% first named

\bibitem{chimera_meaning} M. Panaggio and D. Abrams, Chimera states: Coexistence of coherence and incoherence in networks of coupled oscillators, Nonlinearity {\bf 28}, R67 (2015).

%\bibitem{chimera_DS} D. Abrams and S. Strogatz, Chimera states in a ring of nonlocally coupled oscillators, Int. J. Bifurcation Chaos {\bf16}, 21 (2005).

\bibitem{chimera_AME} A. Zakharova, M. Kapeller, and E. Sch\"{o}ll, Chimera death: Symmetry breaking in dynamical networks, Phys. Rev. Lett. {\bf112}, 154101 (2014).

\bibitem{chimera_DRSD} D. Abrams, R. Mirollo, S. Strogatz, and D. Wiley, Solvable model for chimera states of coupled oscillators, Phys. Rev. Lett. {\bf101}, 084103 (2008).

\bibitem{chimera_GAF} G. Sethia, A. Sen, and F. Ataym, Clustered chimera states in delay-coupled oscillator systems, Phys. Rev. Lett. {\bf100}, 144102 (2008). %%%% time-delay

%\bibitem{chimera_OYP} O. E. Omel'chenko, Y. Maistrenko, and P. Tass, Chimera states: The natural link between coherence and incoherence, Phys. Rev. Lett. {\bf100}, 044105 (2008).

\bibitem{chimera_OMY} O. E. Omel'chenko, M. Wolfrum, and Y. Maistrenko, Chimera states as chaotic spatiotemporal patterns, Phys. Rev. E {\bf81}, 065201(R)  (2010).

\bibitem{chimera_SMK} S. Nkomo, M. tinsley, and K. Showalter, Chimera states in populations of nonlocally coupled chemical oscillators, Phys. Rev. Lett. {\bf110}, 2144102 (2013).

\bibitem{chimera_SR} S. Ujjwal and R. Ramaswamy, Chimeras with multiple coherent regions, Phys. Rev. E {\bf88}, 032902  (2013).

\bibitem{chimera_C} C. Lang, Chimera states in heterogeneous networks, Chaos {\bf 19}, 013113 (2009).

%\bibitem{chimera_TG} T. Ko and G. Ermentrout, Partially locked states in coupled oscillators due to inhomogeneous coupling, Phys. Rev. E {\bf78}, 016203 (2005).

\bibitem{chimera_SY} S. Shima and Y. Kuramoto, Rotating spiral waves with phase-randomized core in nonlocally coupled oscillators, Phys. Rev. E  {\bf69}, 036213 (2004).   %%%%% two-dimensional latices

%%%%%%%%%%%%%  different oscillating dynamics
\bibitem{maps1} I. Omelchenko, Y. Maistrenko, P. H\"{o}vel, and E. Sch\"{o}ll, Loss of coherence in dynamical networks: Spatial chaos and chimera states, Phys. Rev. Lett. {\bf106}, 234102 (2011).

\bibitem{maps2} I. Omelchenko, B. Riemenschneider, P. H\"{o}vel, Y. Maistrenko, and E. Sch\"{o}ll, Transition from spatial coherence to incoherence in coupled chaotic systems, Phys. Rev. E {\bf85}, 026212 (2012).

\bibitem{SL_model} C. Laing, Chimeras in networks of planar oscillators, Phys. Rev. E {\bf81}, 066221 (2010).

\bibitem{HR} J. Hizanidis, V. Kanas, A. Bezerianos, and T. Bountis, Chimera states in networks of nonlocally coupled Hindmarsh-Rose neuron models, Int. J. Bifurcation Chaos {\bf24}, 1450030 (2014).

%%%%%%%%%%%%%  multi-channel cou-plings
\bibitem{FHN} I. Omelchenko, O. E. Omel'chenko, P. H\"{o}vel, and E. Sch\"{o}ll,
    When nonlocal coupling between oscillators becomes stronger: Patched synchrony or multichimera states, Phys. Rev. Lett. {\bf110}, 224101 (2013). %%%%%% multi-channel cou-plings

%%%%%%%%%%%%%  different network structures
\bibitem{chimera_complex} Y. Zhu, Z. Zheng, and J. Yang, Chimera states on complex networks, Phys. Rev. E {\bf89}, 022914 (2014).  %%%complex network

%%%%%%%%%%%%%  new phenomena of chimera state
\bibitem{transients} M. Wolfrum and O. E. Omel'chenko, Chimera states are chaotic transients, Phys. Rev. E {\bf84}, 015201(R) (2011).

\bibitem{tworegions1} Y. Zhu, Y. Li, M. Zhang, and J. Yang, The oscillating two-cluster chimera state in non-locally coupled phase oscillators, Europhys. Lett. {\bf 97}, 10009 (2012).

%\bibitem{tworegions2} Y. Zhu, M. Zhang, and J. Yang, Reversed two-cluster chimera state in non-locally coupled oscillators with heterogeneous phase lags, Europhys. Lett. {\bf 103}, 10007 (2013).

%%%%%%%%%%%%%  In experiments
\bibitem{chemical} M. Tinsley, S. Nkomo, and K. Showalter, Chimera and phase-cluster states in populations of coupled chemical oscillators, Nat. Phys. {\bf8}, 662 (2012).

\bibitem{electronic} L. Larger, B. Penkovsky, and Y. Maistrenko, Virtual chimera states for delayed-feedback systems, Phys. Rev. Lett. {\bf111}, 054103 (2013).

\bibitem{optical} A. Hagerstrom, T. Murphy, R. Roy, P. H\"{o}vel, I.
    Omelchenko, and E. Sch\"{o}ll, Experimental observation of chimeras in coupled-map lattices, Nat. Phys. {\bf8}, 658 (2012).

%%%%%%%%%%%%%  control chimera

\bibitem{control_JOM} J. Sieber, O. E. Omel'chenko, and M. Wolfrum, Controlling unstable chaos: Stabilizing chimera states by feedback, Phys. Rev. Lett. {\bf112}, 054102 (2014).

\bibitem{control_OYP} O. E. Omel'chenko, Y. Maistrenko, and P. Tass, Chimera states induced by spatially modulated delayed feedback, Phys. Rev. E {\bf82}, 066201 (2010).

\bibitem{control_CE} C. Bick and E. Martens, Controlling chimeras, New J. Phys. {\bf17}, 033030 (2015).

%%%%%%%%%%%%%  cluster synchronization

\bibitem{CS:Hasler} Hasler, Yu. Maistrenko, and O. Popovych, Simple example of partial synchronizaiton of chaotic systems, Phys. Rev. E {\bf 58}, 6843 (1998).

\bibitem{CS:YZHANG} Y. Zhang, G. Hu, H. Cerdeira, S. Chen, T. Braun, and Y. Yao, Partial synchronization and spontaneous spatial ordering in coupled chaotic systems, Phys. Rev. E {\bf 63}, 026211 (2001).

\bibitem{CS:Pikovsky} A. Pikovsky, O. Popovych, and Yu. Maistrenko, Resolving clusters in chaotic ensembles of globally coupled identical oscillators, Phys. Rev. Lett. {\bf 87}, 044102 (2001).

\bibitem{CS:BAO} B. Ao and Z. Zheng, Partial synchronization on complex networks, Europhys. Lett. {\bf 74}, 229 (2006).


%%%%%%%%%%%%%  cluster definition
%\bibitem{cluster_definition} E. Basar, \emph{Brain Function and Oscillation} (Springer, New York, 1998).

%%%%%%%%%%%%%  cluster synchronization studies
\bibitem{cs_ZHOUCS} C. Zhou and J. Kurths, Hierarchical synchronization in complex networks with heterogeneous degrees, Chaos {\bf 16}, 015104 (2006).

\bibitem{CS_WANGEPL} X. G. Wang, S. Guan, Y.-C. Lai, B. Li, and C. H. Lai, Desynchronization and on-off intermittency in complex networks, Europhys. Lett. {\bf 88}, 28001 (2009).

\bibitem{cs_FU1} C. Fu, Z. Deng, L. Huang, and X. G. Wang, Topological control of synchronous patterns in systems of networked chaotic oscillators, Phys. Rev. E {\bf 87}, 032909 (2013).

\bibitem{cs_FU2} C. Fu, W. Lin, L. Huang, and X. G. Wang, Synchronization transition in networked chaotic oscillators: The viewpoint from partial synchronization, Phys. Rev. E {\bf 89}, 052908 (2014).

\bibitem{CS:Pecora2014} L. Pecora, F. Sorrentino, A. Hagerstrom, T. Murphy, and R. Roy, Cluster synchronization and isolated desynchronization in complex networks with symmetries, Nat. Commun. {\bf 5}, 4079 (2014).

%%%%%%%%%%%%%  MSF
\bibitem{MSF1} L. Pecora, and T. Carroll, Master stability functions for synchronized coupled systems, Phys. Rev. Lett. {\bf 80}, 2109 (1998).

\bibitem{MSF2} G. Hu, J. Yang, and W. Liu, Instability and controllability of linearly coupled oscillators: Eigenvalue analysis,  Phys. Rev. E {\bf 58}, 4440 (1998).

\bibitem{MSF3} L. Huang, Q. Chen, Y.-C. Lai, and L. Pecora, Generic behavior of master-stability functions in coupled nonlinear dynamical systems, Phys. Rev. E {\bf 80}, 036204 (2009).

%%%%%%%%% relationship between syn pattern and symmetry

\bibitem{SYM:Dhys} O. D'Huys, R. Vicente, T. Erneux, J. Danckaert, and I. Fischer, Synchronization properties of network motifs: Influence of coupling delay and symmetry, Chaos {\bf 18}, 037116 (2008).

\bibitem{SYM:Russo} G. Russo and J. J. E. Slotine, Symmetries, stability, and control in nonlinear systems and networks, Phys. Rev. E {\bf 84}, 041929 (2011).

\bibitem{SYM:Golubitsky} M. Golubitsky and I. Stewart, Recent advances in symmetric and network dynamics, Chaos {\bf 25}, 097612 (2015).

%%%%%%%%%%%%%  computational group theory
\bibitem{CGT} W. Stein, \emph{SAGE: Software for Algebra an Geometry Experimentation} (http://www.sagemath.org/sage/, 2013).

%%%%%%%%%%%%%  Nepal power-grid
\bibitem{Nepal} Nepal electricity authority annual report 2011 (avaliable at http://www.nea.org.np).

%%%%%%%%%%%%% Weighting scheme of network coupling
\bibitem{NETSYN:MOTTER2005} A. E. Motter, C. S. Zhou, and J. Kurths, Enhancing complex-network synchronization, Europhys. Lett. \textbf{69}, 334 (2005).

\bibitem{NETSYN:XGW2007} X. G. Wang, Y.-C. Lai, and C.-H. Lai, Enhancing synchronization based on complex gradient networks, Phys. Rev. E \textbf{75}, 056205 (2007).

%%%%%%%%%%%% applicaiton to biology
%\bibitem{biology1} R. Andersson, A. Johnston, and A. Fisahn, Dopamine D4 receptor activation increases hippocampal gamma oscillations by enhancing synchronization of fast-spiking interneurons, PLos One {\bf 7}, e40906 (2012).

\bibitem{PIN-1} X. F. Wang and G. R. Chen, Pinning control of scale-free dynamical networks, Physica A {\bf 310}, 521 (2002).

\end{thebibliography}
\end{document}